\documentclass[12pt]{article}

\usepackage[breaklinks]{hyperref}
\usepackage{graphicx}
\usepackage{theorem}
\usepackage{amssymb}
\usepackage{euscript}
\usepackage[T1]{fontenc}
\usepackage{color}
\usepackage{xspace}
\usepackage{fullpage}
\usepackage{picins}

\newtheorem{theorem}{Theorem}[section]

\newtheorem{lemma}[theorem]{Lemma}
{\theorembodyfont{\rm}\newtheorem{defn}[theorem]{Definition}}
   {\theorembodyfont{\rm} \newtheorem{remark}[theorem]{Remark}}

\newenvironment{proof}{{\em Proof:}}{\hfill{\hfill\rule{2mm}{2mm}}}
\newcommand{\cardin}[1]{\left| {#1} \right|}
\newcommand{\pth}[2][\!]{#1\left({#2}\right)}
\newcommand{\floor}[1]{\left\lfloor {#1} \right\rfloor}
\newcommand{\permut}[1]{\left\langle {#1} \right\rangle}
\renewcommand{\Re}{{\rm I\!\hspace{-0.025em} R}}

\newcommand{\lemlab}[1]{\label{lemma:#1}}
\newcommand{\lemref}[1]{Lemma~\ref{lemma:#1}}
\newcommand{\figlab}[1]{\label{fig:#1}}
\newcommand{\figref}[1]{Figure~\ref{fig:#1}}

\newcommand{\etal}{\textit{et~al.}\xspace}

\newcommand{\atgen}{\symbol{'100}}
\newcommand{\SarielThanks}[1]{\thanks{Department of Computer
      Science;
      University of Illinois;
      201 N. Goodwin Avenue;
      Urbana, IL, 61801, USA;
      {\tt sariel\atgen{}uiuc.edu}; {\tt
         \url{http://www.uiuc.edu/\string~sariel/}.} #1}}
\definecolor{blue25}{rgb}{0,0,0.25}
\newcommand{\emphic}[2]{%
     \textcolor{blue25}{%
         \textbf{\emph{#1}}}%
         \index{#2}}
\newcommand{\emphi}[1]{\emphic{#1}{#1}}

\newcommand{\PntSet}{{\mathsf{P}}}
\newcommand{\QTree}{\EuScript{T}}
\newcommand{\Cell}{\Box}
\newcommand{\QT}{\mathcal{QT}}
\newcommand{\pnt}{\mathsf{p}}
\newcommand{\num}{z}
\newcommand{\Grid}{\mathsf{G}}

\newcommand{\MakeBig}{\rule[-.2cm]{0cm}{0.4cm}}
\newcommand{\sep}[1]{\,\left|\, {#1} \MakeBig\right.}
\newcommand{\pbrcx}[1]{\left[ {#1} \right]}
\newcommand{\Prob}[1]{\mathop{\mathbf{Pr}}\!\pbrcx{#1}}

\newcommand{\Ex}[1]{\mathop{\mathbf{E}}\!\pbrcx{#1}}

\newcommand{\brc}[1]{\left\{ {#1} \right\}}
\newcommand{\cl}{\mathrm{c{}l}}

\begin{document}

\title{Randomized Incremental Construction of Compressed Quadtrees}

\author{Sariel Har-Peled\SarielThanks{}}

\date{\today}

\maketitle

\begin{abstract}
    We present a simple randomized incremental algorithm for building
    compressed quadtrees. The resulting algorithm seems to be simpler
    than previously known algorithms for this task.
\end{abstract}

%%%%%%%%%%%%%%%%%%%%%%%%%%%%%%%%%%%%%%%%%%%%%%%%%%%%%%%%%%%%%%%%%%
%%%%%%%%%%%%%%%%%%%%%%%%%%%%%%%%%%%%%%%%%%%%%%%%%%%%%%%%%%%%%%%%%%

\section{Introduction}

In this note, we point out that compressed quadtrees can be built via
randomized incremental construction.  Compressed quadtrees are simple
geometric data-structure. Despite their simplicity, they are
surprisingly useful for carrying out various geometric tasks, see
\cite{h-gaa-08}.

The first randomized algorithm for building compressed quadtrees is
due to Clarkson \cite{c-faann-83}.  Eppstein \etal \cite{egs-sqsdd-05}
suggested building compressed quadtrees by using hierarchical random
sampling in a style similar to skip-lists. If one allows bitwise
operations (in particular, interleaving the bits of two integers in
constant time) one can build compressed quadtrees using $z$-order
\cite{g-ewrq-82, h-gaa-08} by a relatively simple algorithm, but the
task is more challenging if such operations are not allowed.

The new algorithm we describe seems to be quite simple, and can be
interpreted as a variant of the skip quadtree of Eppstein \etal
\cite{egs-sqsdd-05}.

\section{Preliminaries}

\begin{defn}[Grid.]
    For a real positive number $\num$ and a point $p = (x, y)$ in
    $\Re^2$, define $\Grid_\num(p)$ to be the grid point
    $\pth[]{\floor{x/\num} \num, \floor{y/\num} \num}$.  Observe that
    $\Grid_\num$ partitions the plane into square regions, which we
    call grid \emphic{cells}{cell}. Formally, for any $i,j \in
    \mathbb{Z}$, the intersection of the half-planes $x \geq \num i$,
    $x < \num(i+1)$, $y \geq \num j$ and $y < \num(j+1)$ is said to be
    a grid \emphi{cell}.
\end{defn}

\begin{defn}[Canonical square.]
    A square is a \emphi{canonical square}, if it is contained inside
    the unit square, it is a cell in a grid $\Grid_r$, and $r$ is a
    power of two.  
\end{defn}

Given a set $\PntSet$ of $n$ points in the unit square, a quadtree
$\QTree$ is built as follows: The root corresponds to the unit square.
Every node $v \in \QTree$ corresponds to a cell $\Cell_v$ (i.e., a
square), and it has four children. The four children correspond to the
four squares formed by splitting $\Cell_v$ into four equal size
squares, by horizontal and vertical cuts. The construction is
recursive, and we start from $v = \mathrm{root}_\QTree$. As long as
the current node contains more than, say, two points of $\PntSet$, we
create its children, and continue recursively the construction in each
child. We stop when each leaf of this tree contains a single point of
$\PntSet$.

\parpic[r]{\begin{minipage}{7.6cm}
    \includegraphics{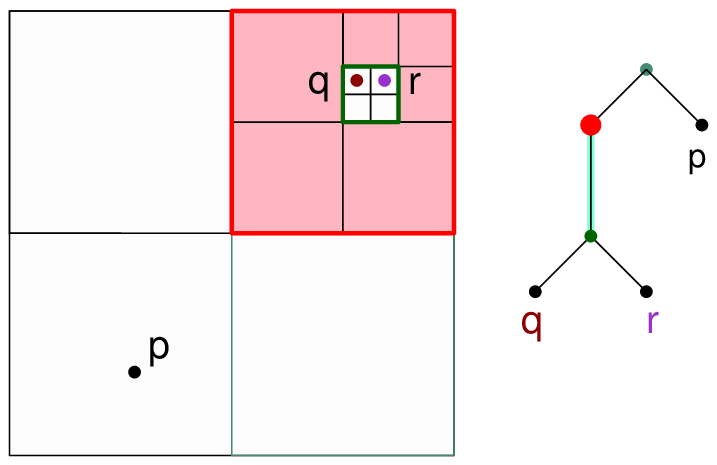}
    \caption{A compressed edge corresponds to a tile that is the set
       difference of two canonical squares.}
    \figlab{tile}
    \medskip
\end{minipage}
}

By compressing paths (in this tree) of nodes that all have a single
child, we get a compressed quadtree of size $O(n)$. Let $\QT(\PntSet)$
denote the (uniquely defined) \emphi{compressed quadtree} of
$\PntSet$.

A leaf in this quadtree corresponds to a canonical square, and a
compressed edge (or more precisely the top vertex of this edge)
corresponds to an annulus formed by the set difference of two
canonical squares. We will refer to such region as a \emphi{tile}, see
\figref{tile}; that is, a tile is either a square (corresponding to a
leaf of the compressed quadtree) or an annulus (corresponding to a
compressed edge). As such, a compressed quadtree induces a partition
of the unit square into these tiles. We denote the planar map induced
by these tiles of the compressed quadtree of $\PntSet$ by
$\QT(\PntSet)$.

\section{Algorithm and analysis}

\subsection{The algorithm}

Pick a random permutation $\permut{\PntSet} = \permut{\pnt_1, \ldots,
   \pnt_n}$ of the points of $\PntSet$. Let $\QTree_i$ be the
compressed quadtree of $\PntSet_i = \brc{\pnt_1, \ldots, \pnt_i}$. 
% We
%denote the corresponding permutation of $\PntSet_i$ by
%$\permut{\PntSet_i} = \permut{\pnt_1, \ldots, \pnt_i}$. 
In any node of $\QTree_i$ that corresponds to a tile $f$ of
$\QT(\PntSet_i)$, we store a list, denoted by $\cl(f)$, of all the
points of $\PntSet$ that lie inside $f$. As such, any point of
$\PntSet$ is stored exactly once somewhere in $\QTree_i$. We will
refer to $\cl(f)$ as the \emphi{conflict list} of $f$.  We also store
for every point of $\PntSet$ a pointer to the node of $\QTree_i$ that
contains it.

In the $i$th iteration, we find the node $v_i$ of $\QTree_{i-1} =
\QT(\PntSet_{i-1})$ that stores $\pnt_i$, and we insert $\pnt_i$ into
this node.  This insertion might result in at most a constant number
(i.e., three) of new nodes being created.\footnote{Here is a sketch
   why this claim is correct: Only a leaf of a compressed quadtree
   might contain an inserted point. As such, we might need to
   introduce a new leaf to store the new point $\PntSet_i$. Hanging
   this new leaf in tree might require splitting an existing
   compressed edge of $\QTree_{i-1}$, by introducing a new
   vertex. Similarly, if the leaf $f$ we insert $\pnt_i$ into already
   stores an inserted point, then we need to introduce a new leaf not
   only for $\pnt_i$ but also for the previously stored point in this
   leaf. This might also result in a new compressed edge if two points
   are close together compared to the diameter of $f$.}  The resulting
tree $\QTree_{i}$ is the compressed quadtree of $\PntSet_i$.  Now, we
need to move all the points stored in $v_i$ to their new proper place
in $\QTree_{i}$.  Thus, for every point stored in $v_i$, we check if
it has to now be stored in one of the new nodes, and if so we move it
to this new node.  If there are $k$ points in the conflict list of
$v_i$ then this iteration takes $O(1 + k)$ time.

The compressed quadtree $\QTree_n$ is the required tree.

\subsection{The analysis}

\begin{defn}
    Let $Y$ be an arbitrary subset of $\PntSet$, and consider a tile
    $f \in \QT(Y)$. A set $X \subseteq Y$ is a \emphi{defining set}
    for $f$, if $f \in \QT(X)$ and it is a minimal set with this
    property (i.e., no proper subset of $X$ has $f$ as a tile).
\end{defn}

The following is proved by a tedious but easy case analysis.
\begin{lemma}
    If $X$ is a defining set of a tile $f \in \QT(\PntSet)$ then
    $\cardin{X} \leq 4$.

    \lemlab{tedious}
\end{lemma}

Unlike ``traditional'' randomized incremental construction, the
defining set is not unique in this case.

\begin{lemma}
    Consider a tile $f \in \QT(\PntSet_i)$. The probability that $f$
    was created in the $i$th iteration is $\leq 4/i$. Formally, we
    claim that
    \[
    \Prob{ f \in \QT(\PntSet_i) \setminus  \QT(\PntSet_{i-1})
       \sep{ f \in \QT(\PntSet_i)}} \leq \frac{4}{i}.
    \]

    \lemlab{backward}
\end{lemma}

\begin{proof}
    Let $D_1, \ldots, D_m \subseteq \PntSet_i$ be all the different
    defining sets of $f$. Consider the set $Z = D_1 \cap D_2 \cap
    \cdots \cap D_m$. 
    
    Observe that $f$ was created in the $i$th iteration only if
    $\pnt_i \in Z$. Indeed, if $\pnt_i \notin Z$, then there exists a
    defining set $D_t$ of $f$ such that $\pnt_i \notin D_t$. But then,
    $f$ is also a tile of $\QT(\PntSet_{i-1})$ as $D_t \subseteq
    \PntSet_{i-1}$, and the probability of this tile to be created in
    the $i$th iteration is zero.

    Now, by \lemref{tedious}, all the defining sets have cardinality
    at most four, and $\cardin{Z} \leq 4$. As such, the required
    probability is bounded by the probability that $\pnt_i$ is in $Z$.
    We bound this probability by backward analysis.  Indeed, fix the
    set $\PntSet_i$ and consider all possible permutations of this
    set. The probability that one of the (at most) four points of $Z$
    is the last point in this permutation (of $i$ elements) is at most
    $4/i$.
\end{proof}

\medskip

Observe that the probability of a tile $f$ to be created (according to
\lemref{backward}) is independent of the size of its conflict list.

\begin{lemma}
    The expected amount of work in the $i$th iteration is $O(1 +
    n/i)$.

    \lemlab{iteration}
\end{lemma}

\begin{proof}
    Consider a tile $f \in \QT(\PntSet_i)$. The amount of work spent
    on it, if it was created in the $i$th iteration, is proportional
    to the size of its conflict list $\mathrm{cl}(f)$.  Let $X_i$ be
    the random variable which is the amount of work spend by the
    algorithm in the $i$th iteration.  Since the total size of the
    conflict lists of $\QTree_i$ is $n$, we get by \lemref{backward} that the
    expected work in the $i$th iteration is bounded by
    \[
    \Ex{ X_i \sep{ \PntSet_i}} = O \pth{ 1 + \sum_{f \in
          \QT(\PntSet_i)} \frac{4}{i} \cardin{\cl(f)} } = O\pth{ 1 +
       \frac{n}{i}}.
    \]
    (Again, the expectation here is over all possible permutations of
    $\PntSet_i$.)  Now, we have that $\Ex{X_i} = \Ex{\Ex{ X_i \sep{
             \PntSet_i}}} = O(1+n/i)$.
%    \aftermathA
\end{proof}

\begin{theorem}
    Given a point set $\PntSet$ of $n$ points in the plane contained
    inside the unit square, one can build a compressed quadtree for
    $\PntSet$ in $O(n \log n)$ expected time.
\end{theorem}
\begin{proof}
    By \lemref{iteration}, the total expected work of the above
    algorithm is $O \pth{ \sum_{i=1}^n \pth{1 + n/i}} = O(n \log n)$.
\end{proof}

\begin{remark}
    The algorithm can also be analyzed using the results from Clarkson
    \etal \cite{cms-frric-93}.
\end{remark}
%\bigskip

%------------------------------------------------------------------
%------------------------------------------------------------------

\section{Discussion and conclusions}

The algorithm presented for building quadtrees works also for points
in higher dimensions.

It is natural to compare our algorithm to Eppstein \etal
\cite{egs-sqsdd-05}. They get a slightly more complicated algorithm,
but they support both insertions and deletions, while our algorithm
can only build the quadtree. In light of our approach, it is natural
to interpret the algorithm of Eppstein \etal \cite{egs-sqsdd-05} as a
lazy randomized incremental algorithm for building quadtrees
\cite{bds-lric-95}. 

The author believes that this is a neat example of backward
analysis. The reader naturally has the right to disagree.

\section*{Acknowledgments}

The author thanks Ken Clarkson and David Eppstein for useful
discussions on the problem studied in this note.

%-------------------------------------------------------------------------
 
\bibliographystyle{alpha} 
\bibliography{shortcuts,geometry}

\end{document}